# Laser damage thresholds of ITER mirror materials and first results on in situ laser cleaning of stainless steel mirrors

M Wisse[1], L Marot[1], B Eren[1], R Steiner[1], D Mathys[2] and E Meyer[1]

[1] Department of Physics, University of Basel, Klingelbergstrasse 82, CH-4056 Basel, Switzerland

[2] Centre of Microscopy, University of Basel, Klingelbergstrasse 50/70, CH-4056, Basel, Switzerland

Email: marco.wisse@unibas.ch, tel: +41 61 267 37 25, fax: +41 61 267 37 84

**Abstract**

A laser ablation system has been constructed and used to determine the damage threshold of stainless steel, rhodium and single-, poly- and nanocrystalline molybdenum in vacuum, at a number of wavelengths between 220 and 1064 nm using 5 ns pulses. All materials show an increase of the damage threshold with decreasing wavelength below 400 nm. Tests in a nitrogen atmosphere showed a decrease of the damage threshold by a factor of two to three. Cleaning tests have been performed in vacuum on stainless steel samples after applying mixed Al/W/C/D coatings using magnetron sputtering. In situ XPS analysis during the cleaning process as well ex situ reflectivity measurements demonstrate near complete removal of the coating and a substantial recovery of the reflectivity. The first results also show that the reflectivity obtained through cleaning at 532 nm may be further increased by additional exposure to UV light, in this case 230 nm, an effect which is attributed to the removal of tungsten dust from the surface.

## 1. Introduction

In ITER, the most crucial element of each optical system is the first mirror in the optical path, which serves to guide light into the mirror labyrinth leading to the diagnostic located outside the neutron shielding. Depending on their exact location in the machine, the first mirrors may be subject to erosion and/or deposition, both of which will affect the reflectivity of the mirror surfaces and compromise the measurements. Though finding a suitable material is still an ongoing effort [1,2], the most promising candidates at present are molybdenum and rhodium, on account of their reflectivity and resilience to sputtering. In particular their single crystalline form has demonstrated good performance compared to the more common (and cheaper) polycrystalline form. However, the prohibitive cost and the difficulty of manufacturing large size single crystalline mirrors have motivated the investigation of nanocrystalline coatings, such as may be produced by magnetron sputtering, as an alternative [3,4,5].

Whatever the final choice of material will be, an issue that is still outstanding is how to prevent the degradation of the optical performance due to the build-up of deposits containing beryllium, tungsten, deuterium, tritium and possibly carbon, depending on the final choice of divertor material. This is a concern not only for mirrors in deposition dominated areas, but also in erosion dominated areas, as even in the latter the surface may be buried under a deposit. The rate of formation of deposits, their exact elemental composition and the resulting degradation of the reflectivity are difficult to predict, and have motivated the investigation of mitigation techniques.

In some cases the mirror may be protected by a shutter, a gas buffer, or by placing it inside a structure that is especially designed to reduce the incident particle flux. However, these measures cannot be expected to be either 100% effective or applicable in all cases. Therefore, a way must be found to remove deposits in situ, in the presence of a magnetic field of 0.8 or even 5 Tesla, depending on whether cleaning takes place during plasma operation or in between pulses, where the toroidal field could be decreased.

The two cleaning methods that are currently being considered are plasma cleaning and laser cleaning, both of which have their own merits and disadvantages. Plasma cleaning might be implemented in the form of a local RF or DC generated plasma to erode the deposit either chemically or physically through sputtering [6]. In principle such a method could operate continuously, thereby preventing a deposit from building up at all [7]. Also, the worldwide industrial expertise in plasma processing inspires confidence that an engineering solution could be found for implementation in ITER. It is, however, difficult to envisage how such a method might work in the presence of a strong magnetic field, which would reduce the contact between the plasma and the mirror surface in the same way that it reduces the contact between the fusion plasma and the wall of the tokamak.

**2. Laser cleaning**
In laser cleaning, pulsed laser radiation is used to remove contaminating layers through ablation or delamination. Laser cleaning has been used successfully for a wide variety of cleaning tasks, such as cleaning artworks, satellite dishes and the removal of paint in hazardous environments. In addition, pulsed lasers have found widespread application in micromachining and ophthalmology, where they are successfully used for surface modification [8].

Although the application of pulsed lasers for the purpose of surface modification has certainly matured for the kind of applications typically found in industry and medicine, it has to be said that this is not the case for the mirror cleaning problem faced by the fusion community. Although a number of experiments have been carried out, experiments to date have been largely confined to the nanosecond regime and to wavelengths accessible by lasers normally used for Thomson scattering. Also, most experiments have been carried out in air. This is, of course, due to financial restrictions. However, it is clear from the literature available on laser ablation that a wider parameter range should be explored to assess the viability of laser cleaning for the purpose of cleaning ITER first mirrors. To justify this claim, we shall briefly explore the physical processes usually assumed to be involved in laser cleaning, as well as summarize the results obtained in the fusion community so far, before describing the first results obtained at the University of Basel.

*2.1. Physical mechanisms of laser cleaning.*
There are a number of processes that can play a role in the removal of material from a surface by pulsed laser radiation, depending on the type of material and the duration, wavelength and fluence of the laser pulse [9-16]. The first step is the absorption of the laser radiation by electrons in the material. In a metal this means absorption by free electrons. In the case of a material with a bandgap, electrons have to be excited into the conduction band before substantial absorption can take place. Population of the conduction band may take place through avalanche ionization and/or multiphoton ionization, depending on the fluence of the laser pulse. The relative roles of avalanche and multiphoton ionization are discussed in [9] and [13]. In avalanche ionization, seed electrons gain energy in the laser field and ionize bound electrons through collisions, producing more free electrons. According to [15], avalanche ionization is responsible for the ablation of wide bandgap materials at laser intensities below $10^{12}$ W/cm$^2$. At intensities above $10^{13}$ W/cm$^2$, multiphoton absorption of bound electrons becomes the dominating mechanism for the excitation of bound electrons into the conduction band. As the density of free electrons increases, the absorption of energy from the laser pulse increases until it reaches a maximum when the electron density is such that the plasma frequency equals the laser frequency, at which point the plasma becomes opaque. For pulses in the sub-picosecond range, the material removal occurs mainly after the pulse, either through the process of thermal evaporation, in which electron-phonon collisions increase the local temperature above the vaporization point, or through Coulomb explosion, whereby escaping electrons create a field strong enough to pull ions out of the surface [15,16]. The former dominates at high laser fluence, whereas the latter dominates at fluences near the ablation threshold. In the case of longer laser pulses, however, in particular those in the nanosecond domain, thermal diffusion during the laser pulse becomes important. In this case, the free electrons have the time to thermally equilibrate with the ion lattice during the pulse, resulting in a much larger heat affected zone than in the case of ultrashort pulses, where the heat affected zone is confined to the optical skin depth. The ablation process is now thermal, whereby the ions escape the surface when their thermal energy exceeds the binding energy. In this regime, the ablation threshold fluence is known to scale with the square root of the pulse duration [10,16], a relation which breaks down for short pulses (~ ps), which have been reported to exhibit a decrease of the threshold fluence slower than this [17].

Important to note is that the electronic structure has little effect on the ablation process in the case of ultrashort pulses, because the required free electrons are produced mainly by multiphoton ionization, whereas longer, lower fluence pulses rely on seed electrons during the initial stage of the absorption process [15]. This means that short pulse ablation is repeatable, whereas longer pulse ablation, at least of non-metals, depends on statistical variations of the free electron density and is inherently less predictable. Furthermore, it has been shown [18] that the ablation threshold is much sharper for short pulses as it is for long pulses, as may be understood intuitively from the different mechanisms involved.

The wavelength dependence of the ablation threshold still appears somewhat controversial. According to [9], the ablation threshold should increase with decreasing wavelength in the case of avalanche ionization. This is contradicted in [13], according to which there should be only a mild dependence. Multiphoton ionization, on the other hand, becomes more efficient at shorter wavelengths and so the damage threshold decreases with decreasing wavelength.

Though the above considerations apply to the removal of bulk material, there are other situations to consider, such as dust and films. Dust removal by laser pulses has been extensively studied for various different types of dust and is known to depend on particle size, shape and the type of material as well as the surface roughness [8]. The angle of incidence also plays a role, though contradictory results have been reported, [19] reporting a higher efficiency for perpendicular incidence and [20,21] claiming the opposite. For strongly absorbing particles it is believed that the sudden absorption of heat and the associated thermal expansion of either or both the dust particles and the substrate impart sufficient momentum to the dust particles to overcome the adhesion force and leave the surface. This is more difficult for smaller particles, as the force per unit area is larger [8]. This mechanism is assumed to be responsible for the removal of carbon dust in [22], as the particle removal efficiency was found to be independent of wavelength. Removing tungsten particles was found to be easier at shorter wavelengths and ultrashort pulses [23], however. To explain this it is proposed that the emission of photoelectrons causes ionization of ambient oxygen, resulting in electrostatic forces pulling the dust off the surface. An improvement with decreasing wavelength was also found with the removal of $SiO_2$ particles [21].

The removal of films is possible through any of the mechanisms associated with ablation, but in addition they may be removed by so-called buckling [8], where local heat absorption leads to thermal stress and buckling of the film, followed by the formation of cracks around the buckled region and the subsequent dislodging of the affected part of the film. In the case of transparent films, the delivery of laser energy to material depends on the film thickness and interference effects may introduce a strong wavelength dependence.

It is likely that the composition, morphology and thickness of the deposits that will form on the ITER first mirrors will vary greatly depending on the location in the machine, as it depends on the local particle densities, their fluxes, the local temperature as well as the current state of the mirror, which, in turn, will depend on its history. It can therefore hardly be expected that a unique laser cleaning configuration can be found that works sufficiently well in all cases. In fact, the insensitivity of the ablation process to the electronic structure in the case of femtosecond pulses would favour those, whereas the dust removal and buckling mechanisms benefit from long pulses, as it is heat transfer to the ion lattice that is key. Also, femtosecond pulse delivery would be even more difficult from the engineering point of view than nanosecond pulse delivery, as femtosecond pulses cannot be delivered using optical fibres due to group velocity dispersion. Furthermore, femtosecond systems tend to be quite bulky and have a low power output compared to longer pulse system.

*2.1. Laser ablation experiments in the fusion community*
In the fusion community, experiments have been performed to clean windows, such as those used for Thomson scattering [24,25], to remove carbon deposits containing tritium [26-28] and to clean mirrors [29-32].

Zhou et al. [29] report the removal of a 1 micron carbon film from a polycrystalline molybdenum film exposed in HL-2A, using a YAG laser at 1064 nm with a pulse length of 10 ns. The molybdenum mirror is reported to have a damage threshold of 1.0 J/cm$^2$, whereas the deposit was found to ablate at 0.6 J/cm$^2$.

Leontyev et al. [30] determined the damage threshold of molybdenum and stainless steel using a high repetition ytterbium rate fibre laser delivering 120 ns pulses at 20 kHz and at 1.06 micron. They report a difference between the damage threshold obtained for exposure to a single pulse and multiple pulses (~5000), indicating that cumulative damage plays a role. For molybdenum, the single and multipulse thresholds are reported to be 6.46 J/cm$^2$ and 3.1 J/cm$^2$ respectively. For stainless steel the single pass threshold is given as 2.26 J/cm2, while the multipulse threshold was not measured. It is interesting to compare the single shot damage threshold of molybdenum to the value obtained by Zhou et al. Though Zhou et al.have not explicitly stated that they measured the single shot threshold, this can be inferred from the text. The ratio of Leontyev's value to Zhou's is 6.46/1.0=6.46, while the ratio of pulse durations is 120/10=12. Assuming the square root dependence of the threshold fluence with laser pulse duration to apply, we should expect the ratio of threshold fluences to have been about 3.5. However, it unclear what type of molybdenum Leontyev used.

In [27], Grisolia et al.report on experiments aimed at detritiation of carbon tiles by ablation of the deposit on the top surface, using a YAG laser delivering 100 ns pulses at 532 nm at 10 kHz. Bare graphite is reported to ablate at 2.5 J/cm$^2$, whereas the deposit ablates at only 0.5 J/cm$^2$. However, with a 5 ns pulse the ablation of pure graphite is found to start at 1 J/cm2. This again contradicts the square root scaling mentioned before, as this would suggest the threshold to drop by a factor of 4.5, not 2.5. Note, however, that this scaling may be somewhat oversimplified. In this case, for example, one might imagine that a pulse as long as 100 ns leads to negligible plasma plume formation and hence to relatively little light loss, whereas a short pulse may suffer a greater loss due to plasma and dust formation. Alternatively, the effect of cumulative damage could be smaller in the case of short pulses.

The first ablation experiments on substrates containing beryllium were performed by Widdowson et al. [31], who investigated the cleaning of stainless steel and molybdenum mirror samples taken from the JET tokamak after two years of exposure. The experiments were performed using a system apparently identical to the one used by Leontyev, i.e. 1064 nm, 120 ns pulses at 20 kHz, who performed his experiments to determine a suitable fluence that could be safely used for the experiments on the JET samples that were to follow. Unfortunately, despite taking these precautions, the mirrors did get damaged in the experiment, though a substantial recovery of the reflectivity was achieved.

Experiments on beryllium surfaces were reported in [32], which describes experiments on a set of single crystalline molybdenum mirrors coated either with a 420 nm carbon or 150 nm beryllium film in the PISCES-B facility, using 1064 nm, 220 ns pulses at 8-10 kHz. The experiments on beryllium were done in an argon atmosphere, while carbon was treated also in vacuum. Though carbon was successfully removed at 1.8 J/cm$^2$, this was not the case for beryllium, even at 2.7 J/cm$^2$. Removing the carbon film proved less effective in an argon atmosphere near atmospheric pressure than in vacuum, which resulted in charring of the exposed carbon. It is postulated that a shorter pulse, of the order of 5 ns, may lead to better results.

In summary, the experiments carried out so far have concentrated on ablating carbon layers, beryllium layers and mixed layers on stainless steel and single- and polycrystalline molybdenum using wavelengths of 532 nm and 1064 nm. Most experiments have been performed in air, though one experiment in vacuum has been carried out. All experiments have been performed in the thermal regime, i.e. with pulse lengths typically longer than the electron-lattice equilibration time, and therefore lead to a large heat affected zone. No experiments have as yet been done to establish whether the ablation threshold is wavelength dependent. Also, no data exists on the ablation of layers containing a mixture of carbon, tungsten and beryllium. The only data on beryllium containing layers to date have been obtained in air, using very long pulses, while it is expected that short pulses are preferable. In addition, two of the candidate materials, rhodium and nanocrystalline molybdenum, have not been investigated at all.

Though the above list is not exhaustive, it gives a fair impression of the parameter range that has been covered to date. Despite the partial recovery of the reflectivity of the samples from JET, the damage incurred by the mirror surface during the cleaning process led to a decline of the interest in pursuing a laser cleaning solution to the ITER mirror problem. Given the limited parameter range explored so far, this decision seems rather premature, in particular in view of the fact that very few other options remain, while the success of ITER crucially depends on the proper functioning of its diagnostics and, in turn, the importance of fusion on the European political agenda depends entirely on the success or failure of the ITER project.

At the University of Basel we have started addressing some of these issues. The present work focuses on the ablation threshold of stainless steel, rhodium, and single-, poly- and nanocrystalline molybdenum in vacuum, and its wavelength dependence. Also, cleaning experiments in vacuum have been performed on stainless steel coated with mixtures of carbon, tungsten and aluminium, the latter being used as a substitute for beryllium. All experiments have been done using 5 ns pulses, though experiments using picosecond pulses are currently being planned. A discussion of the ablation of pure metal layers containing only tungsten and aluminium on polycrystalline molybdenum is deferred to a next publication. This assumes particular relevance since it seems to have been decided that the carbon divertor that was originally planned for the ITER start-up phase will be skipped for financial reasons, so that ITER is now expected to operate with a tungsten divertor from the start.

**3. Plasma deposition and laser ablation setup**
At the University of Basel, a laser ablation system has been added to an existing plasma exposure facility [33]. This facility offers a complete toolkit for the manufacturing and investigation of metallic mirrors, including experimental simulation of erosion and deposition conditions that are likely to occur in ITER. This is supplemented by an extended set of diagnostics for chemical surface analysis and optical surface characterization as well as plasma diagnostic tools, including in situ X-ray and UV photoelectron spectroscopy, ex situ spectroscopic ellipsometry and spectrophotometry as well as scanning electron microscopy (SEM) and energy dispersive X-ray spectroscopy (EDX). The setup, shown in Figure 1, is constructed around a high vacuum chamber equipped with an RF plasma source and magnetrons, allowing the injection of various types of gases and the admixture of impurities through magnetron sputtering. Conventional pumping systems are used to obtain a background pressure of typically $1.5 \times 10^{-4}$ Pa. A surfatron operating at 13.56 MHz and a typical power of 90 W is used to create a plasma in a Pyrex tube of 12 cm diameter and 40 cm length. At the heart of the chamber is a sample carousel with heatable and biasable probe holders, which typically take samples of 25 mm diameter. A photoelectron spectrometer (Leybold EA 10 MCD, Mg anode) is attached to the vacuum chamber, allowing samples to be transferred directly for chemical surface analysis after exposure.

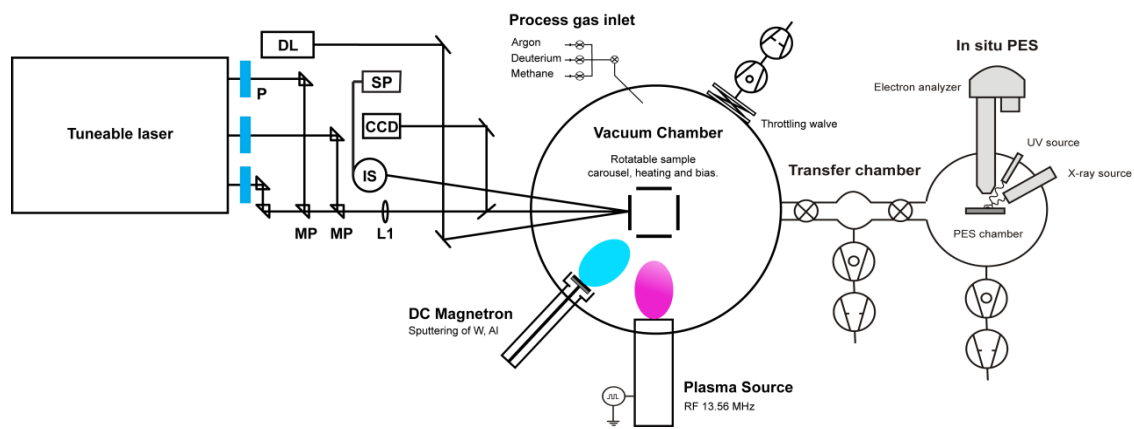

*Figure 1: Schematic of the plasma exposure facility with added laser ablation system and in situ photo-electron spectrometer. Tuneable pulsed laser radiation (210-2300 nm, 20 Hz, 5 ns pulses) is emitted from one of three ports and attenuated by filters and/or polarisers (P), guided along the same beam path by moveable prisms (MP) and focused onto the sample with a plano-convex lens (L1). A diode laser (DL) beam is coupled into the chamber to monitor the evolution of the reflectivity as well as to keep track of the position of the main laser beam. The reflected beam is captured by an integrating sphere (IS) coupled to a USB spectrometer (SP)*

The laser is a commercial optical parametric oscillator (OPO) system (EKSPLA NT 342B) delivering tuneable, polarized laser radiation between 210 and 2300 nm at 20 Hz, with a 5ns pulse length and pulse energies between 1.5 mJ and 35 mJ depending on wavelength. The output beam is round, with a diameter of 6 mm and exits the laser through one of three ports, depending on wavelength. The beam divergence is less than 2 mrad. The pulse energy may be attenuated either by optical filters or by rotating BBO polarizers. The beams from each of the three ports are guided along the same beam path by two movable prisms, mounted on computer controlled translation stages (Thorlabs PT1/M-Z8). The beam is then focused onto the sample through a quartz viewport with a plano-convex quartz lens element, a typical spot size being of the order of 0.5 mm. A CCD beam profiler (Dataray WinCam D UCD15) is used to measure the beam profile both in the visible and in the UV. For measurements in the UV, a UV convertor (Dataray BSF12G12N-AR1) is used, which employs a fluorescent material to convert the UV light into visible radiation for detection by the CCD. The beam is sent to the beam profiler by a moveable mirror that may be inserted after the focusing lens, the distance between the lens and the profiler being the same as the distance between the lens and the sample. A laser energy meter (Gentec SOLO 2 / QE12P-S-MB) is used to measure the pulse energy. Measurement of the beam profile together with the pulse energy allows the calculation of the energy density at the beam focus, which may be up to 5 J/cm$^2$. The evolution of the reflectivity of the sample spot being exposed to the OPO beam is monitored using a diode laser (Edinburgh instruments EPL 475). To this end, the diode beam is focused onto the same spot and is captured using an integrating sphere connected to a USB spectrometer (Avantes Avaspec 2048-USB2-UA). This way, the measurement is insensitive to small changes in the orientation of the diode beam when scanning the sample. The sample holder, which may be heated by a boron nitride heater, is equipped with computer controlled xy-translation stages (Zaber NA08A25 vacuum, with KT-MCA controller) that serve to scan the laser beam across the sample. Two chromel/alumel thermocouples monitor the temperature of the sample and the motor housing. Care has been taken to minimise the thermal conductivity between the heated part of the probe holder and the translation stages, resulting in a negligible rise of the temperature of the latter when heating the sample.

## 4. Laser damage threshold measurements

In order to prevent damage to the mirror surface during cleaning it is necessary to establish the maximum fluence the material can withstand without melting, as a function of wavelength. Also, it is desirable to know whether the atmosphere has an influence on the damage threshold.

Tests have been performed on five different types of material, including stainless steel (316L), rhodium coated on stainless steel [4], single crystalline molybdenum (110), polycrystalline molybdenum and nanocrystalline molybdenum coated on stainless steel by magnetron sputtering [3]. The samples were of the order of 20-25 mm and either round or square depending on the material.

The procedure was as follows. The sample was mounted in the vacuum chamber, moved to the starting position and heated to 150 degrees C. A wavelength was chosen and the beam profile measured. Then, the pulse energy was adjusted to the desired value by a combination of optical filters and a polariser, starting with a high value so that the damage could be easily seen and the alignment adjusted if necessary. A series of shots were then fired while monitoring the change of the reflectivity with the diode laser, after which the beam was moved to the next spot by moving the sample and the pulse energy was decreased. This procedure was repeated until no change of the reflectivity was observed after exposure. The number of shots fired at each particular spot varied depending on the material, as it turned out that in some cases damage occurred only after a substantial number of pulses had been fired. The minimum number of shots fired was 1000, corresponding to 50 seconds of exposure at 20 Hz, while the maximum was 8000, corresponding to 400 seconds of exposure. A picture of a sample being exposed is shown in Figure 2.

The nature of the damage induced by laser radiation depends on the energy density. Power levels in the range of several J/cm$^2$ typically result in local melting, as shown in Figure 2a and b. However, focusing of the laser light by microparticles may cause crater formation at lower power levels as well. Depending on the atmosphere, chemical reactions such as oxidation may occur at the surface, resulting in coloured annuli surrounding the impact region (Figure 2c). Even when the power density is not high enough to cause such drastic changes to the surface, another effect that is typically observed is the exposure of the grain boundaries and the presence of ripples with a period of the order of the wavelength. (Figure 2d, electronic version). These so-called laser induced periodic structures are well known and have been investigated in detail by various authors [34-38].

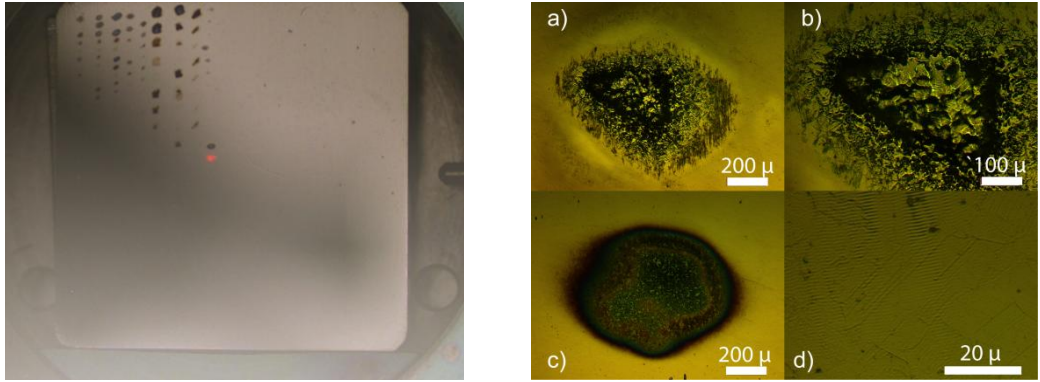

*Figure 2: A 25 x 25 mm sample during laser damage threshold measurements. The columns of spots are where the sample has been exposed at successively lower energies while moving down the column, changing wavelength after 3-5 spots. (a)-(d): Typical damage observed when exceeding the damage threshold include melting (a,b) and discolouration (c), which occurs in the presence of reactive gases like oxygen or nitrogen. In (d), the grain boundaries have become visible and the areas between them show ripples with a period of the order of the laser wavelength.*

In order to deduce the energy density, the beam profile measurements are combined with energy meter readings. Typical beam profiles of the OPO and the diode laser are shown in Figure 3. It is clear that the spot produced by the OPO is much larger than the one from the diode, which is a natural consequence of the much higher beam quality of the diode as compared to the OPO, the former being Gaussian and the latter being more tophat shaped. As both the OPO beam and the diode laser beam have a finite size and their maxima are unlikely to coincide exactly, simply taking the maximum value of the OPO profile would lead to an overestimate of the energy density sampled by the diode beam. As a conservative approach we convolve the measured diode profile with the OPO profile use the maximum value of this profile instead.

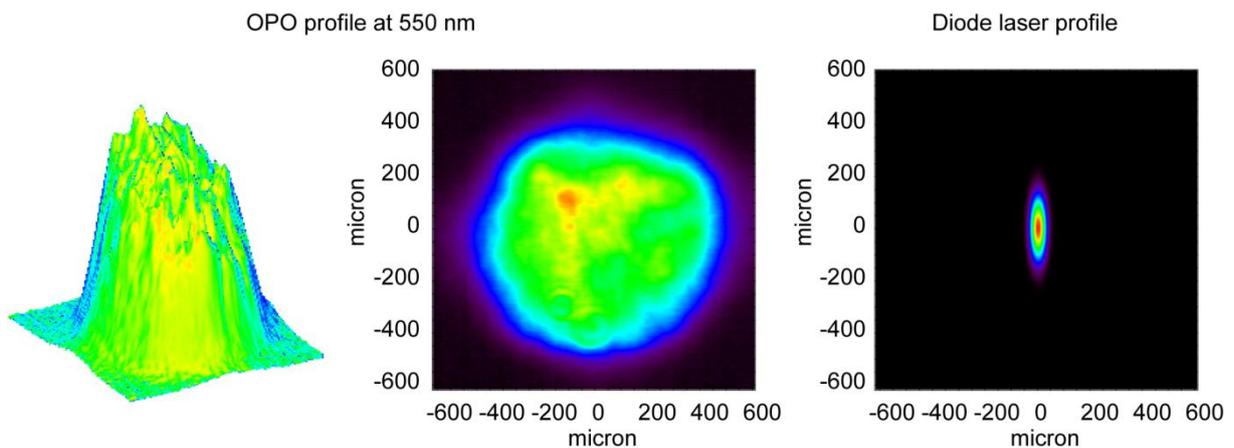

*Figure 3: Profiles of the OPO beam at 550 nm and the diode beam at the beam focus. The OPO beam is round and tophat-ish, while the diode laser beam is elliptic and Gaussian.*

The results of measurements on the five materials mentioned above, for wavelengths between 220 and 1064 nm at 50 nm intervals are shown in Figure 4. The interesting thing to note is that all samples show an increasing damage threshold with decreasing wavelength in the UV. As avalanche photoionization should be the dominating mechanism for the ablation of metals in the nanosecond regime, this trend is in keeping with the suggestion that the threshold fluence should increase with decreasing wavelength [9].

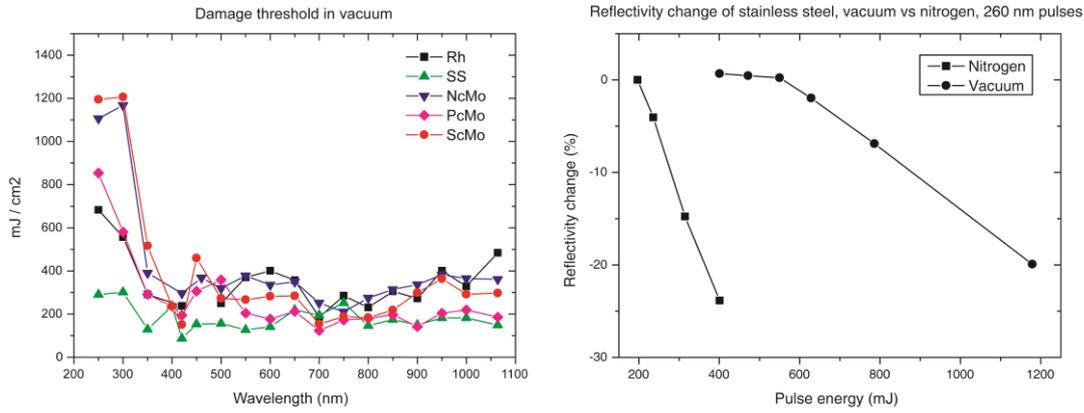

*Figure 4: (left) Damage thresholds as a function of wavelength for different materials. Note that both nanocrystalline and single crystalline molybdenum perform quite well. All materials show an increasing resilience to damage in the UV. (right) Change of the reflectivity of stainless steel after 1000 pulses at 260 nm as a function of pulse energy, in vacuum and in the presence of nitrogen.*

In order to facilitate comparison with the findings of other authors we chose to express the threshold fluence in units of pulse energy per unit area. The multipulse damage threshold for molybdenum at 1064 nm reported by Zhou [29] of 1 J/cm$^2$ for 10 ns pulses is rather higher than the value of the order of 0.3 J/cm$^2$ we find. Similarly, we find that stainless steel gets damaged at about 0.2 J/cm$^2$, whereas the value of 2.3 J/cm$^2$ for 120 ns pulses at 1064 nm found by Leontyev [30] would lead us to expect a threshold of around 0.4 J/cm$^2$ for 5 ns pulses. The fact that our experiments were carried out in vacuum means that the discrepancy is even greater, as we established that the threshold fluence is larger in vacuum. Hence, if we had carried out our experiments in air, like Zhou and Leontyev, we would have found even smaller values. There are many possible causes of this discrepancy, including the method used to establish the occurrence of damage, the exact nature of the material, the method to calculate the laser beam energy distribution, the pulse energy measurement etc. We are, therefore, not too concerned by this difference and would rather focus on the scaling of the threshold fluence with wavelength and the difference between vacuum and air.

*4.1. Vacuum vs. nitrogen atmosphere*
Another question was whether the presence of air would have an effect on the damage threshold. In order to investigate this, the procedure described above was repeated on stainless steel for a number of wavelengths, with and without flooding the vacuum chamber with nitrogen. Figure 4 shows the results obtained at 260 nm. The difference is quite dramatic, showing an increase of the damage threshold by about a factor of two to three in vacuum. Similar results were obtained at 300, 400 and 532 nm. Irradiation above the threshold typically resulted in the formation of coloured annuli. Though the formation of chemically reactive species might play a role in a nitrogen atmosphere, a similar difference in the ablation threshold between vacuum and air was observed by Gamaly et al. [39], who reported a difference of over a factor of two for the ablation of aluminium, copper, steel and lead using 12 ps, 532 nm pulses. The model they propose to explain these results does not assume the creation of chemically active species during the laser pulse, but suggests that the formation of a Maxwellian distribution at the surface, necessary for thermal ablation to play a role, is accelerated in the presence of air. In vacuum, on the other hand, thermal ablation would be negligible at this pulse length. Therefore, the presence of air offers an additional mechanism to contribute to material removal, leading to a lower ablation threshold. Note that this effect should not play a role for pulses in the femtosecond regime, which has indeed been confirmed in [10].

## 5. Cleaning tests on stainless steel

Although stainless steel is not being considered for application as a mirror surface in ITER, it is, of the five materials tested here, the one that is most easily damaged by laser radiation. Therefore, it is the most interesting material to subject to initial cleaning tests, as the ability to clean it without damaging would inspire confidence that the other materials can be cleaned successfully as well.

*5.1. Deposition of ITER relevant contaminating layers*

The elemental composition of the deposits that will form in ITER is difficult to ascertain and will depend strongly on the location in the machine. According to modelling it will be a mixture of beryllium, tungsten, deuterium, tritium and, at the time of writing, possibly carbon, depending on the final decision on the divertor material [40,41]. Due to the toxicity of beryllium and the severe restrictions on handling this material, consensus that was also adopted for this study is to use aluminium as a beryllium substitute, on account of its chemical properties and its expected behaviour in mixed layers such as those that might be expected to occur in ITER. Due to its radioactivity, tritium is equally difficult (and undesirable) to handle. However, it is chemically very similar to deuterium and using only deuterium is perfectly adequate.

Using two pulsed DC magnetrons, the RF surfatron described earlier, deuterium as a working gas and 1% by partial pressure of methane, layers containing different relative amounts of aluminium tungsten, carbon and oxygen were created on stainless steel mirror samples, each with a thickness of the order of a few hundred nanometers. Before coating, the samples were cleaned for 10 minutes in pure deuterium plasma with a bias of -200 V. It should be noted that in all cases but one, the sample temperature was floating during the experiment, resulting in a temperature of about 50 degrees C. Table 1 shows the elemental composition of the four coatings, determined from XPS measurements of the Al 2p, W 4f, C 1s and O 1s lines. Note that XPS cannot be used to measure deuterium. From other studies using similar films [42] it is known that the elements are present in the form of tungsten carbide and aluminium oxide. In this study we will only present the relative amounts of each element.

*Table 1: Elemental composition of the layers deposited on each of the mirror samples. [1]) RF deuterium plasma preseent during deposition. [2]) Sample heated to 150 degrees C during deposition.*

| Sample | % Al | % W | % C | % O |
|---|---|---|---|---|
| A[1] | 31 | 7 | 20 | 42 |
| B | 16 | 32 | 50 | 2 |
| C | 39 | 1 | 39 | 21 |
| D[1,2] | 15 | 36 | 39 | 10 |

*5.1. Laser cleaning procedure*

Instead of moving the laser beam, the beam is scanned across the sample by moving the sample, which is mounted on a pair of computer controlled translation stages. To clean a rectangular area, the sample performs a scanning motion in the xy-direction. The speed depends on the spot size of the beam and the desired overlap between consecutive pulses, taking into account that the laser fires at 20 Hz. The spot size of the beam chosen for the cleaning tests described below was 300 micron, with an overlap of 250 micron, meaning that the sample travelled 50 microns every 50 milliseconds, i.e. with 1 mm per second. The overlap exists for both the x- and y-direction, so that in this case the sample moved by 50 micron in the y-direction after each scan line in the x-direction. The dimensions of the cleaned areas were typically 17 x 17 mm, resulting in a total time of about 1.5 hours per cleaning cycle.

A number of different experiments were performed on the various samples. In the first experiment, a large area on sample A was cleaned at 250 nm to allow XPS and reflectivity measurements. Then, two overlapping patches on sample B were exposed at different wavelengths. Finally, samples C and D were cleaned completely, using the knowledge obtained in the previous experiments.

## 6. Results

*6.1. Sample A*
An area of 17 x 17 mm was cleaned using 250 nm, 1 mJ pulses, which is slightly above the damage threshold found earlier. Figure 5 shows a picture of the sample as well as three SEM images of the surface after cleaning. The coating has been removed, but the surface appears milky and the SEM images reveal what appears to be dust. Also, small circular holes can be seen.

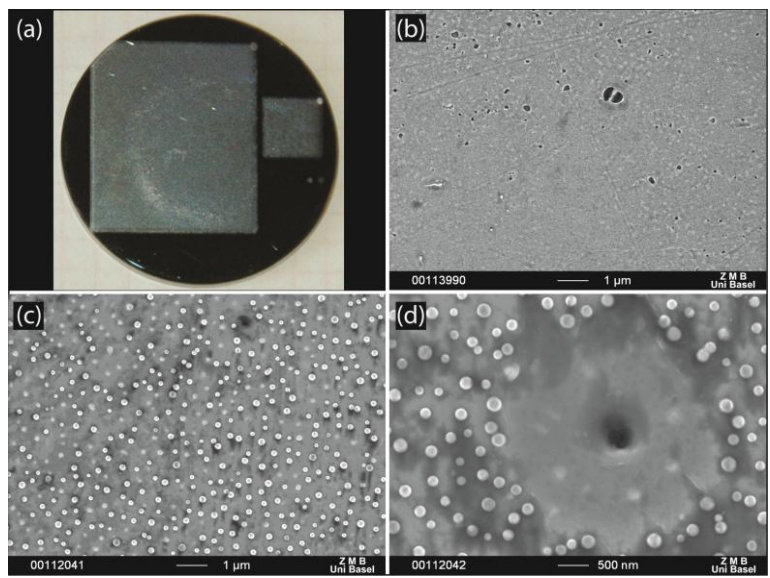

*Figure 5: (a) Image of sample A, showing two cleaned patches and the remaining coating. (b) SEM image of the coating. (c) SEM image of the central cleaned patch, showing fine tungsten dust. (d) Tungsten dust around a crater.*

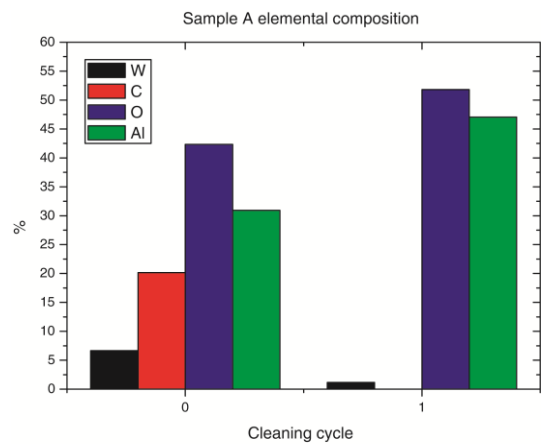

*Figure 6: Elemental composition of sample A before and after cleaning at 230 nm. Carbon has disappeared completely and the amount of tungsten has decreased, leaving mainly aluminium oxide.*

Figure 6 shows the elemental composition of the surface determined with XPS, which provides no spatial resolution due to the large beam size (~8mm), but is sensitive only to the first few nm of the surface. After cleaning, no carbon is visible and the tungsten concentration has decreased considerably, leaving mainly aluminium oxide.

To determine the composition of the spherical particles left on the surface after cleaning, EDX (Energy Dispersive X-ray spectroscopy) mapping of the surface was performed, shown in Figure 7. Each panel shows the distribution of one particular element. The black spots in the panels corresponding to iron and chromium show the absence of these elements at the location of the particles, showing that the electron beam did not penetrate the dust particles. The bottom right panel identifies the dust as tungsten. The top right panel, showing the distribution of aluminium, shows accumulation of aluminium around the tungsten particles. The spherical shape of the tungsten particles suggests that they were formed through melting during the ablation process. Note that the relative concentration of tungsten as measured by XPS is very small, which is due to the fact that the tungsten dust, as seen by the spectrometer, has a very small surface area, wheras aluminium is sprinkled over the entire surface.

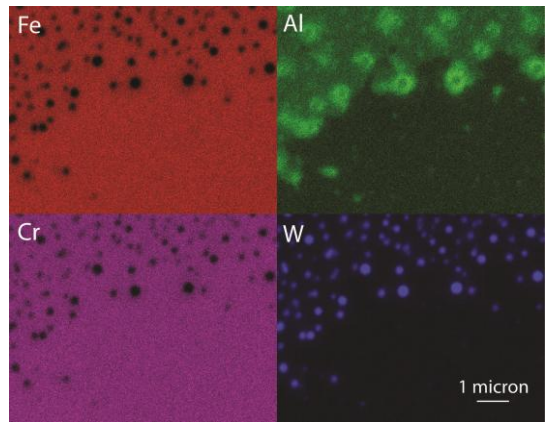

*Figure 7: EDX images of sample A, showing the spatial concentration of Fe, Al, Cr and W, identifying the dust remaining on the surface as W.*

Figure 8 shows the specular and diffuse reflectivity between 250 and 2300 nm, measured with a Varian Carian 5 spectrophotometer before and after coating, and after cleaning. Although a reasonable fraction of the original reflectivity is recovered, the diffuse reflectivity increases considerably, probably due to the tungsten dust remaining on the surface.

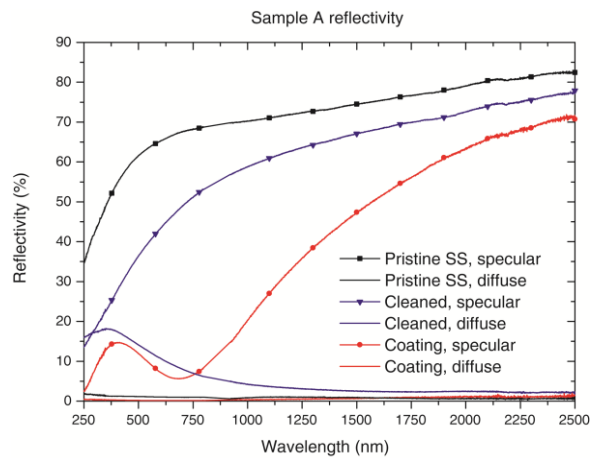

*Figure 8: Specular (symbols) and diffuse (smooth) reflectivity of sample A before coating (black), after coating (red) and after cleaning (blue).*

*6.1. Sample B*

On this sample, two partly overlapping quadratic patches were exposed four times. The first patch was cleaned at 532 nm, 1 mJ/pulse and the second patch at 230 nm, 0.7 mJ/pulse, see Figure 9. The interesting thing to note is the much higher brightness of the overlapping region compared to the non overlapping region. Visual inspection during the cleaning process had shown a sudden increase of the reflectivity already as a result of the first overlapping exposure, suggesting that the increase may not be due just to the additional exposure itself, but rather to an enhanced cleaning effect effected by the succession of two different wavelengths.

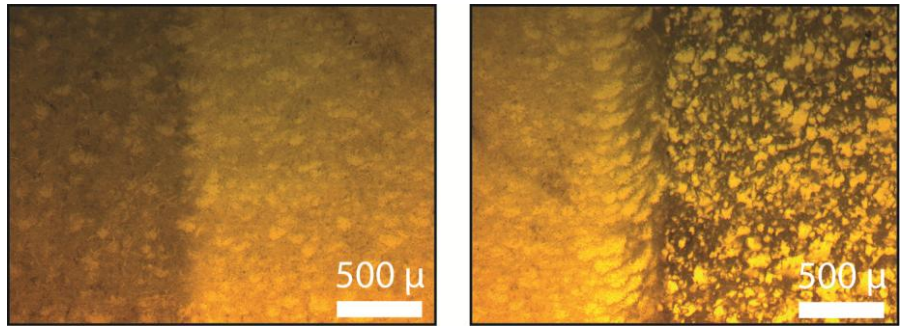

*Figure 9: Two patches were cleaned four times at 532 nm / 1 mJ and 230 nm / 0.7 mJ, corresponding to the dark regions in the left and right figure respectively. The bright region is where the two patches overlap.*

*6.1. Sample C*

In order to investigate the suggestion that a succession of exposures at different wavelengths might lead to an enhanced cleaning effect compared to using a single wavelength, two samples were cleaned in this manner, using the same power as before. Figure 10 shows a picture of the sample after cleaning as well as SEM images of the coating and the cleaned area.

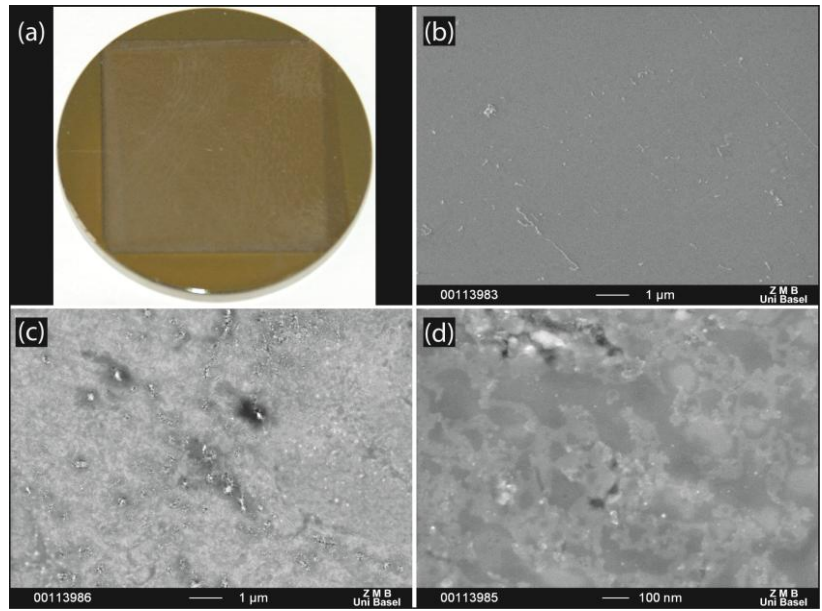

*Figure 10: (a) Sample C after cleaning. (b) SEM image of the coating. (c,d) SEM images of the cleaned area.*

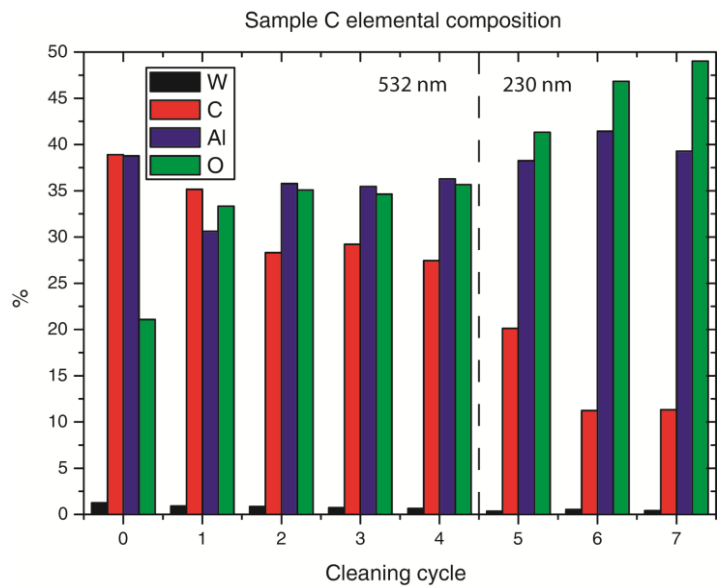

*Figure 11: Surface elemental composition of sample C, determined by XPS measurements, after each cleaning cycle. The relative concentrations remain very similar throughout the exposures at 532 nm. The relative amount of carbon decreases during exposure to 230 nm, while the relative concentration of oxygen increases.*

The evolution of the surface elemental composition was measured using XPS, and is shown in Figure 11, showing the relative atomic concentrations of tungsten, aluminium, carbon and oxygen after each exposure. It may be seen that the relative amount of carbon had stabilized after a number of exposures at 532 nm, but decreased after additional exposure at 230 nm, while the relative oxygen concentration increases. This is most likely to the decreasing thickness of the film, which exposes the oxidized substrate surface.

The EDX spectra taken after exposure are shown in Figure 12. Before and after exposure, iron is the main component in the spectrum and serves as the normalization, showing the decrease of carbon, tungsten and aluminium after exposure. As with sample A, tungsten dust was also formed on this sample, albeit much less. This is reflected also in the smaller increase of the diffuse reflectivity compared to sample A, as shown in Figure 13, which also shows a better recovery of the reflectivity.

As with sample A, dust was observed to form on the surface during the first few cleaning cycles at 532 nm, which was partly removed during irradiation with 230 nm light. The EDX measurement again shows the dust to consists of tungsten. Though the higher efficiency of tungsten dust removal at shorter wavelengths was reported in [22,23], the experiment was carried out in air and the suggestion that ionization of ambient oxygen molecules creates an electric field that pulls the tungsten particles away from the surface obviously does not hold in vacuum. Bearing in mind that the tungsten will be in an oxidized form, the fact that shorter wavelengths and shorter pulses enhance the removal efficiency might suggest that multiphoton absorption plays a role, which would be the case if the particles were being ablated rather than lifted off the surface.

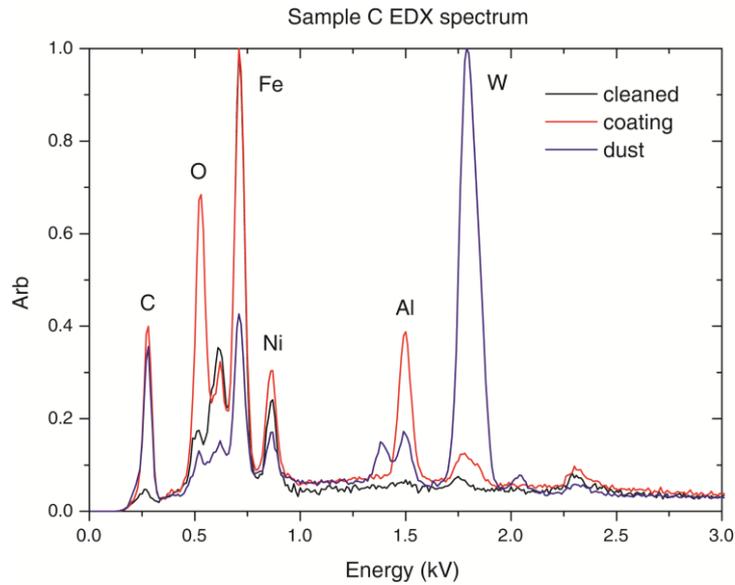

*Figure 12: EDX spectrum of sample C, showing the relative elemental composition before (red) and after (black) cleaning. The blue line corresponds to dust and as before, identifies it as tungsten.*

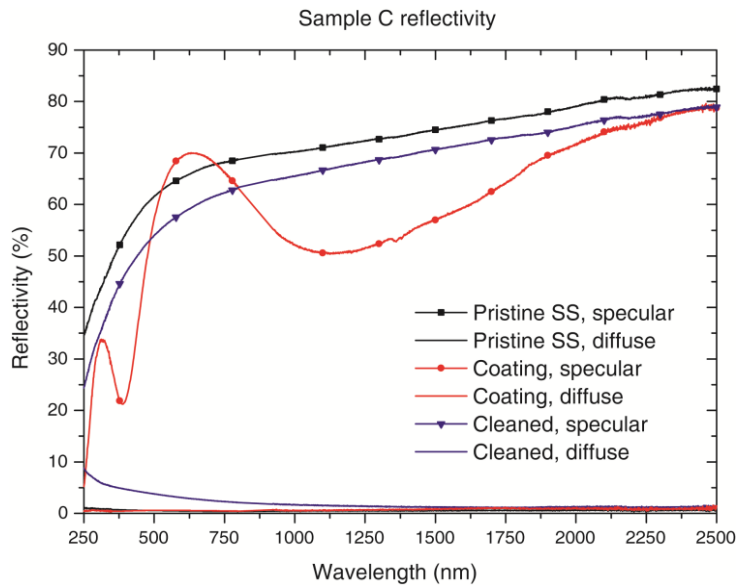

*Figure 13: Specular (symbols) and diffuse (smooth) reflectivity of sample C before coating (black), after coating (red) and after cleaning (blue). The structure of the red line shows that the coating is transparent and causes interference effects.*

6.1. Sample D

The final example had a highly reflective coating, as may be seen from Figure 14 and 16. This sample was exposed twice at 532 nm, 1.3 mJ/pulse, followed by three exposures at 230 nm, 0.5 mJ/pulse.

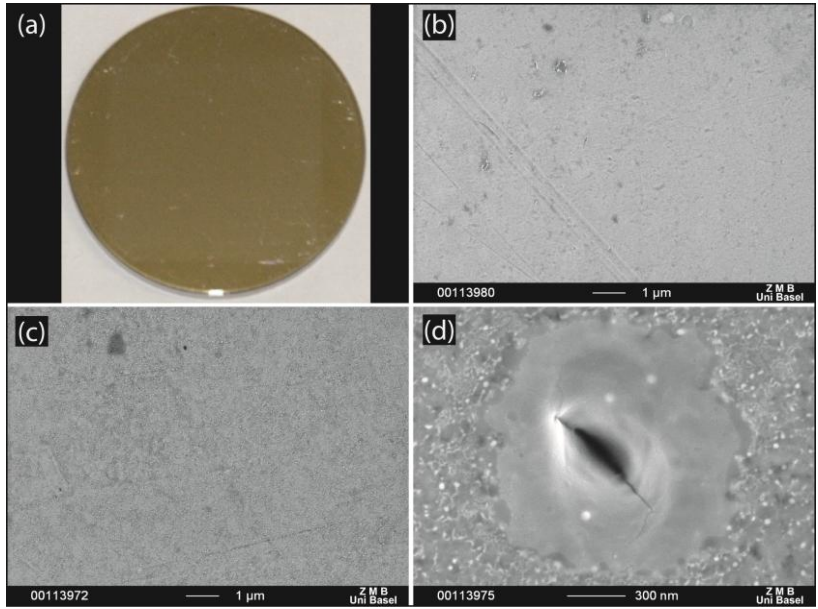

*Figure 14: (a) Picture of sample D after cleaning. The low contrast between the cleaned, square central area and the coating around it illustrates the weak reduction of the reflectivity due to the coating. (b) SEM image of the coating. (c,d) SEM images of the cleaned area, showing fine dust left on the surface (c) and melting (d).*

The evolution of the elemental composition as determined by XPS is shown in Figure 15. EDX data (not shown here) only showed iron features in this case, indicating that the coating was extremely thin. As with sample C, the relative oxygen concentration increases upon exposure at 230 nm, which is attributed to the thinning of the coating, while the carbon content decreases. It is not clear whether the tungsten concentration had stabilized during exposure at 532 nm, but it does continue to drop during the subsequent exposure at 230 nm.

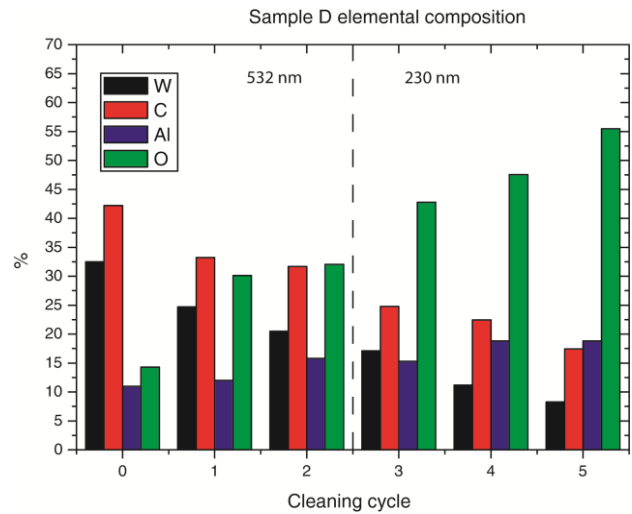

*Figure 15: Elemental composition of sample D after each exposure.*

Figure 16 shows the specular and diffuse reflectivity before an after exposure. As mentioned before, the reflectivity drop due to the coating was relatively small compared to the previous samples. Nonetheless, the cleaning procedure resulted in recovery of the reflectivity comparable to sample C.

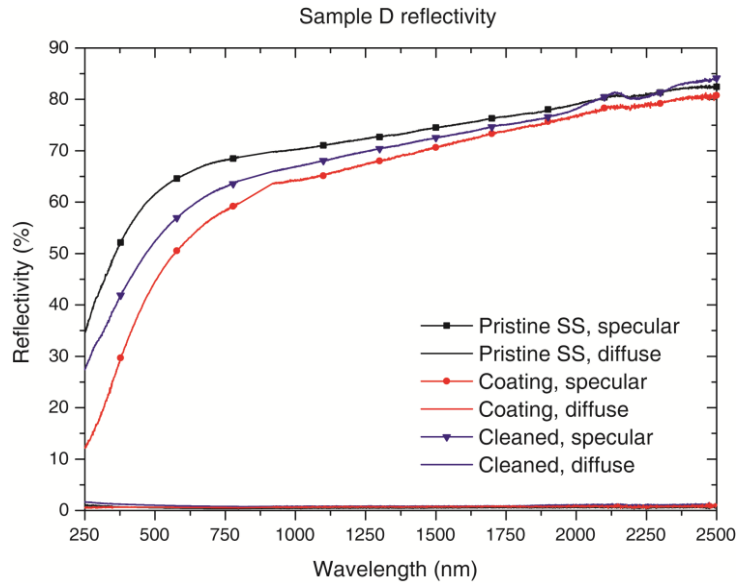

*Figure 16: Specular (symbols) and diffuse (smooth) reflectivity of sample D before coating (black), after coating (red) and after cleaning (blue).*

## 7. Conclusions

A laser ablation system has been successfully developed and integrated with a plasma exposure facility. The system allows the investigation of the complete mirror lifecycle, from the construction of the mirror surface and the subsequent degradation using depositing plasmas to the removal of the deposit using laser cleaning.

Damage thresholds of different ITER relevant mirror materials, including stainless steel, rhodium and single, poly- and nanocrystalline molybdenum have been determined in vacuum, for a range of wavelengths between 220 and 1064 nm. The threshold fluence for sustaining optical damage has been found to be inversely proportional to the wavelength, for wavelengths below 400 nm, for each material. This is consistent with avalanche photoionization being the dominant mechanism in the ablation process.

The damage threshold has been found to be lower in nitrogen than it is in vacuum by about a factor of two to three. Besides the possibility of chemically active species produced by the laser beam playing a role, the explanation given by Gamaly et al. [39] for 12 ps pulses may also apply.

First cleaning tests have been performed in vacuum, on stainless steel with various layers containing different amounts of aluminium, tungsten and carbon. A substantial recovery of the reflectivity has been obtained for each mirror, despite substantial differences in the composition of the coating. The samples were irradiated a number of times using 532 nm, 5 ns pulses at 20 Hz, followed by exposure with 230 nm pulses. Results so far indicate that UV irradiation following visible irradiation can significantly enhance the cleaning efficiency compared to cleaning using a single wavelength. The reason for this is yet to be elucidated, but a few possibilities come to mind. Each sample showed the formation of tungsten dust particles, which may be more efficiently removed at a shorter wavelength as reported in literature, though the removal mechanism remains to be identified in this case. Another possibility, which would apply to the ablation of the particles rather than to their escaping the surface, is that they are in an oxidized form, in which case ablation should become more efficient at shorter wavelengths, as multiphoton ionization would start playing a role in providing electrons to the conduction band to feed the avalanche ionization process. Carbon was successfully removed from each sample. These results are encouraging and motivate further research, as summarized in the following section.

**8. Future work**

As mentioned before, it is hard to predict the nature of the deposits in ITER. Also, it is possible that the properties of the deposits will change during the ablation process, for example through oxidation or dust formation. This provides a strong motivation for the exploration of a much wider parameter space than has been covered to date. One might also argue that the mirror material should be chosen on the basis of being resilient to laser irradiation, in order to facilitate laser cleaning. From that point of view, our results indicate that rhodium and nanocrystalline molybdenum would be a good choice, in particular if UV irradiation should be applied.

Experiments in the short-pulse regime seem called for, given the potential benefits of achieving a smaller heat affected zone as well as a more reproducible and selective ablation process. A collaboration is currently underway to perform experiments in the picosecond regime, which will provide an invaluable addition to the data currently available.

It is very likely that there will be no carbon components in ITER. Therefore, experiments on mixed aluminium/tungsten layers on molybdenum are being carried out and will be reported on in a future publication.

The next experimental campaign will see experiments on the cleaning of polycrystalline molybdenum probes retrieved from the JET vessel. As these probes are contaminated with beryllium and tritium, a special setup has been built to be able to handle these probes. Also included are three polycrystalline mirror samples, coated in PISECES-B with Be.

References

[1] A Litnovsky *et al.* 2007 *J. Nucl. Mater* **363-365** 1395
[2] A Litnovsky *et al.* 2008 *Fusion Eng. Des.* **83** 79-89
[3] B Eren *et al.* 2011 *Fusion. Eng. Des.* **86** 2593-6
[4] L Marot *et al.* 2008 *Surf. Coat. Technol.* **202** 2837
[5] M Joanny *et al.* 2012 *IEEE Transactions on Plasma Science* **40** (30) 692
[6] E Mukhin *et al.* 2009 *Nucl. Fusion* **49** 085032
[7] E E Mukhin *et al.* 2012 *Nucl. Fusion* **52** 013017
[8] Luk'yanchuk, *Laser cleaning vol I and II* (World Scientific Publishing, Singapore, 2002).
[9] A Vaidyanathan *et al.* 1980 *IEEE journal of Quantum Electronics* **16** 89
[10] M D Shirk *et al.* 1998 *J. Laser App.* **10** (1)
[11] B C Stuart *et al.* 1996 *Phys. Rev. B* **53** (4)
[12] P P Pronko *et al.* 1998 *Phys.l Rev. B* **58** (5)
[13] An-Chun Tien *et al.* 1999 *Physical Review Letters* **82** (19)
[14] L Jiang *et al.* 2005 *International Journal of Heat and Mass Transfer* **48** 487-499
[15] L Jiang *et al.*, *Femtosecond lasers ablation: challenges and opportunities*, presented at NSF Workshop on Research Needs in Thermal Aspects of Material Removal, Stillwater, Oklahoma, 2003 (unpublished).
[16] E G Gamalay *et al.* 2002 *Physics of Plasmas* **9** (3)
[17] B C Stuart *et al.* 1995 *Phys. Rev. Lett.* **74** (12)

[18] B C Stuart *et al.* 1996 **13** (2) *J. Opt. Soc. Am. B*

[19] Y W zheng *et al.* 2001 *J. Appl. Phys.* **90** (1) 59

[20] C Curran *et al.*, *Effect of wavelength and incident angle in the laser removal of particles from silicon wafers*, presented at 20th International Congress on Applications of Lasers and Electro-Optics, Jacksonville, 2001 (unpublished).

[21] Y W Zheng *et al.* 2001 *J.Appl. Phys.* **90** (5) 2135

[22] Ph Delaporte *et al.*, *Why using laser for dust removal from tokamaks*, 2010 (unpublished).

[23] A Vatry *et al.* 2011 *J. Nucl. Mater* **415** S1115-S1118

[24] B W Brown *et al.* 1995 *Rev. Sci. Inst.* **66** 4

[25] A Alfier *et al.* 2008 *Rev. Sci. Inst* **79** 10F338

[26] A Widdowson *et al.* 2007 *Am. Nucl. Soc.* 51-54

[27] C Grisolia *et al.* 2007 *J. Nucl. Mater.* **363-365** 1138-1147

[28] A Semerok *et al.* 2007 *J. Appl. Phys* **101** 084916

[29] Y Zhou *et al.* 2011 *J. Nucl. Mater.* **415** S1206-1209

[30] A Leontyev *et al.* 2011 *Fusion Eng.Des.* **86** 1728-1731

[31] A Widdowson *et al.* 2011 *J. Nucl. Mater.* **415** S1199-S1201

[32] C H Skinner *et al.* 2012 *Rev. Sci. Inst.* **83** 10D512

[33] M Wisse *et al.* 2012 *Rev. Sci. Inst.* **83** 013509

[34] L k Ang *et al.* 1998 *J.App. Phys.* **83** (8)

[35] Y F Lu *et al.* 1996 *J. Appl. Phys.* **80** 7052

[36] A Siegman *et al.* 1986 *IEEE Journal of Quantum Electronics* **QE22** (8) 1384

[37] B Tan *et al.* 2006 *J. Micromech. Microeng.* **16** 1-6

[38] J E Sipe *et al.* 1983 *Phys. Rev. B* **27** (2)

[39] E G Gamaly *et al.* 2005 *Phys. Rev. B* **71** 174405

[40] K Schmid *et al.* 2011 *J. Nucl. Mater.* **415** S284-S288

[41] V Kotov *et al.* 2009 *J. Nucl. Mater.* **390-391** 528-531

[42] B Eren *et al.* 2012 *Submitted to J. Nucl. Mater.*

[43] R P Doerner *et al.* 2009 *Nucl. Fusion* **49** 035002

[44] G De Temmerman *et al.* 2007 *J. Nucl. Mater* **363-365** 259-263